\documentclass[pre,twocolumn,amsmath,amssymb,showpacs,superscriptaddress]{revtex4}
\usepackage{graphicx}

\begin{document}
\title{Random field Ising model and community structure in complex networks}

\author{Seung-Woo \surname{Son}}
\affiliation{Department of Physics, Korea Advanced Institute of
Science and Technology, Daejeon 305-701, Korea}
\author{Hawoong \surname{Jeong}}
\affiliation{Department of Physics, Korea
Advanced Institute of Science and Technology, Daejeon 305-701, Korea}
\author{Jae Dong \surname{Noh}}
\thanks{Corresponding author [noh@cnu.ac.kr]}
\affiliation{Department of Physics, Chungnam National University,
Daejeon 305-764, Korea}

\begin{abstract}
We propose a method to find out the community structure of a complex
network. In this method the ground state problem of a ferromagnetic random 
field Ising model is considered on the network with the magnetic field
$B_s = +\infty$, $B_{t} = -\infty$, and $B_{i\neq s,t}=0$ for a
node pair $s$ and $t$.  The ground state problem is equivalent to the
so-called maximum flow problem, which can be solved exactly numerically 
with the help of a combinatorial optimization algorithm.
The community structure is then identified from the ground state 
Ising spin domains for all pairs of $s$ and $t$. 
Our method provides a criterion for the existence of the community
structure, and is applicable to unweighted and weighted networks equally
well. We demonstrate the performance of the method by applying it to the
Barab\'asi-Albert network, Zachary karate club network, the scientific
collaboration network, and the stock price correlation network.
\end{abstract}

\maketitle

\section{INTRODUCTION}
The network theory is a useful tool for the study of complex
systems. Universal features of some biological, social, and
technological systems have been studied through their network
structure~\cite{Albert02,Dorogovtsev02,Newman03R}. Recent studies
have revealed that some complex networks have the community
structure, which means that highly interconnected nodes are clustered 
in distinct parts. The community may represent functional modules in
biological networks~\cite{Jeong00, Holme03, Ravasz03, Wilkinson04}, 
industrial sectors in economic networks~~\cite{Mantegna00, Onnela03},
and cliques of intimate individuals in social 
networks~\cite{Girvan02}.

Recently various methods have been suggested for finding out the
community structure in a given network~\cite{Newman04a}. Girvan
and Newman proposed an algorithm based on iterative removal of
links with the highest betweenness centrality~\cite{Girvan02,
Newman04a, Newman04}. The betweenness centrality of an edge is
given by the number of the pathways passing through it among shortest
paths between all node pairs~\cite{Newman01}.  Nodes in different
communities, if any, would be connected through rare
inter-community links. Hence one could isolate communities by
removing links with the highest betweenness centrality repeatedly.
Similar methods were also considered in Refs.~\cite{Tyler03,
Radicchi04, Fortunato04}. Optimization techniques were also
considered to find out the community structure. In those
approaches, the community structure is found by optimizing an
auxiliary quantity, such as the modularity~\cite{Newman04b,Clauset04}. 
Some physical problems
turned out to be useful in detecting the community structure. For
example, the $q$-state Potts model~\cite{Reichardt04}, the random
walks~\cite{Zhou03}, and the electric circuit
problem~\cite{Wu03} were studied.

Those methods proved to be successful in detecting existing communities.
On the other hand, it would be desirable to develop a
method which can not only detect the community structure but also 
verify its existence.
Most algorithms developed are suitable for unweighted networks,
whereas many real-world networks of interest are weighted~\cite{Newman04c}.
One may modify and generalize the algorithms developed for unweighted
networks. However, such a generalization may not be
straightforward~\cite{Newman04c}. So it would also be desirable 
to develop a method that works for unweighted and weighted networks 
equally well.

In this paper we propose a method for finding out the community
structure, which fulfills the requirements described
above. Our approach is motivated from the observation on the
Zachary network, a classical example of social networks with the
community structure~\cite{Girvan02}. 
It is an acquaintance network of 34 members in a karate club. Once there 
arose a conflict between two influential members, which resulted in the
breakup of the club into two. It is reasonable to think that the members
would tend to minimize the number of broken ties, which can be accomplished
by the breakup in accordance with the community structure.
In fact, the resulting shape after the breakup coincides with the community
structure of the original karate club network~\cite{Girvan02}.
It suggests that the community structure of a given network may be found
by simulating the breakup caused by an enforced frustration among nodes.

We simulate the breakup by studying the ferromagnetic random 
field Ising model~(FRFIM): The Ising spins $\sigma_i=\pm 1$ are assigned to
all nodes $i=1,\ldots,N$, they interact ferromagnetically through links, 
and the quenched random magnetic field $B_i$ is applied to each spin. 
The ferromagnetic interaction represents the cost for broken ties, and the
random field is to introduce the frustration.
In particular, we consider the case where the positive infinite
magnetic field is applied to one spin and the negative infinite
magnetic field to another. It amounts to imposing the boundary
condition that the two spins are in the opposite state. It simulates 
the conflict as raised by the two members in the Zachary network. 
Then, we will identify the community structure from the ground
state spin domain pattern of the FRFIM.

This paper is organized as follows. In Sec.~\ref{sec:2} we introduce
the FRFIM in general weighted networks.
The ground state problem of the FRFIM can be solved exactly with a
numerical algorithm, which will be explained in Appendix.
Then the method for finding out the community structure is presented.
In Sec. \ref{sec:3}, we apply the method to several networks and present
the results.
We conclude the paper with summary and discussion in Sec.~\ref{conclusions}.

\section{METHOD}\label{sec:2}
Consider a weighted network ${G}$ of $N$ nodes.
Connectivity of $G$ can be represented with the weight matrix
$\{J_{ij} | i,j=1,\cdots,N\}$,
where $J_{ij}$ is a prescribed weight or strength of a link between nodes 
$i$ and $j$ if they are connected or $J_{ij}=0$ otherwise.
We assume that the weights are non-negative, $J_{ij}\ge 0$,
and that the weights are symmetric, $J_{ij}=J_{ji}$.
For an unweighted network, the matrix elements take the binary value
$0$ or $1$, and the weight matrix reduces to the usual adjacency matrix.

The FRFIM on the network is defined with the Hamiltonian
\begin{equation} \label{eq:H}
H = - \frac{1}{2}\sum_{i, j} J_{ij} \sigma_i \sigma_j - \sum_{i}
B_i \sigma_i \ ,
\end{equation}
where $\sigma_i = \pm 1$ is the Ising spin variable at each node $i$.
The spins interact ferromagnetically with the coupling strength
$\{J_{ij}\}$. They are also coupled with the quenched random magnetic
field $\{B_i\}$. 

The FRFIM model has been studied extensively in $d$
dimensional regular lattices in order to investigate
the nature of the glass phase transition~(see Ref.~\cite{Middleton02}
and references therein). It has also been studied to investigate the
disorder-driven roughening transition of interfaces in disordered
media~\cite{Noh_Rieger}.
The phase transition in the FRFIM on complex networks
would also be interesting, which has not been studied so far. The issue will
be studied elsewhere~\cite{Son04}.

The specific feature of the FRFIM depends on the distribution
of the random field $\{B_i\}$.
In this work, we consider the simple yet informative
magnetic filed distribution given by
\begin{equation}\label{eq:Bst}
B_i = \left\{
 \begin{array}{cl}
   +\infty & \mbox{, for\ } i=s \\
   -\infty & \mbox{, for\ } i=t \\
  0        & \mbox{, for\ } i \neq s,t
 \end{array} \right.
\end{equation}
for certain two nodes $s$ and $t$. It amounts to imposing the
boundary condition that $\sigma_s = +1$ and $\sigma_t = -1$,
which induces frustration among nodes. This specific random field
distribution is adopted in order to mimic the conflict as in the Zachary
network. In the ground state nodes are separated into different spin domains,
which will be related to the community structure of the underlying network.

\begin{figure}[t]
\includegraphics[width=\columnwidth]{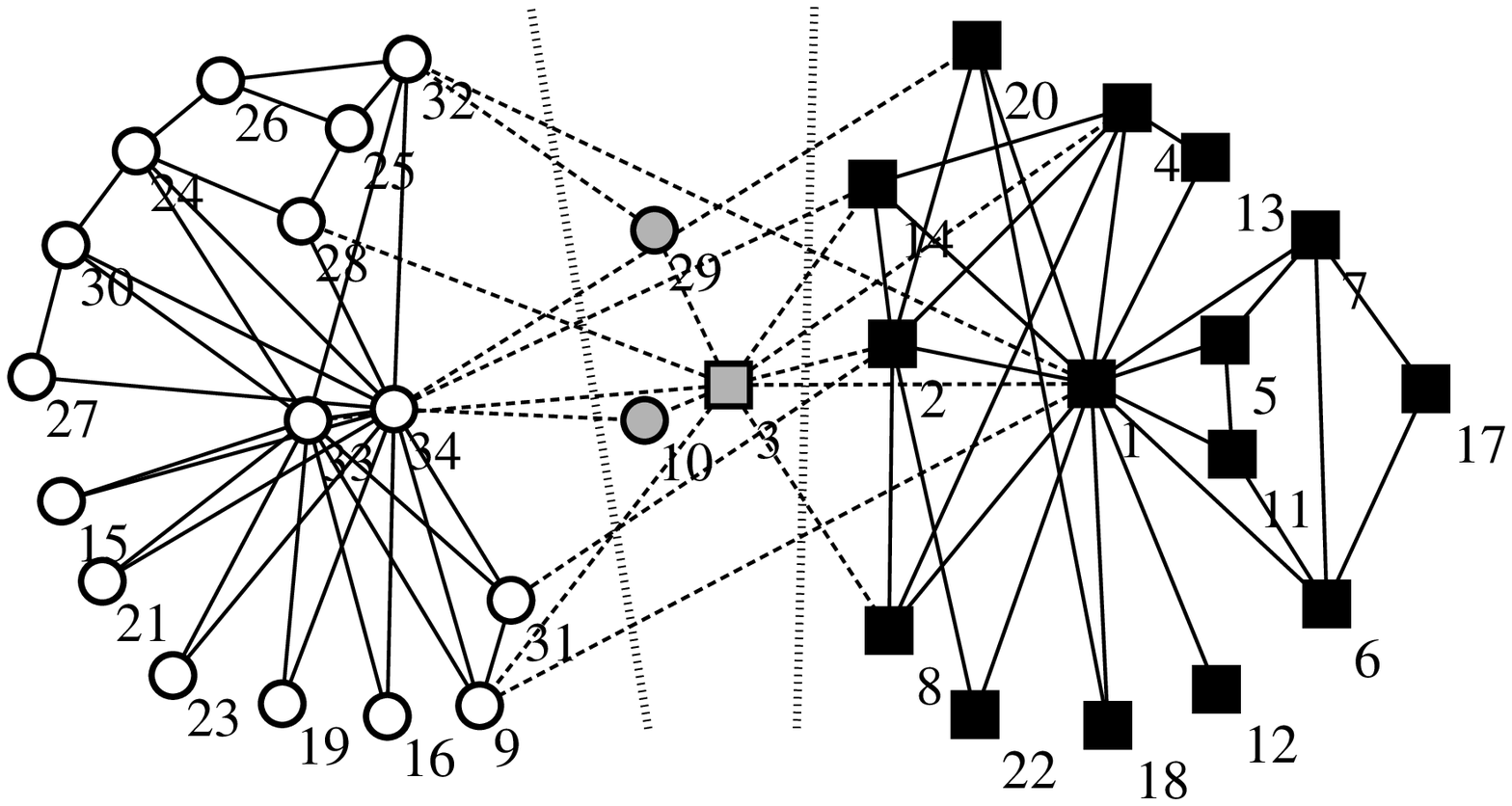}
\caption{Zachary karate club network. The links connecting nodes that are
(not) in the same community are represented with solid~(dashed) lines. The
dotted lines separate the communities.} \label{fig:zachari}
\end{figure}
As an explicit example, we consider the Zachary karate club network
which is illustrated in Fig.~\ref{fig:zachari}. The node labeled
as $1$~($34$) corresponds to the club instructor~(administrator). 
They had a conflict, which resulted in the breakup. 
Nodes in the side of the administrator
and the instructor after the breakup are denoted with circular
and rectangular symbols, respectively. 
With $J_{ij}=1$ for all links and the magnetic field given by 
Eq.~(\ref{eq:Bst}) with $s=1$ and $t=34$, one can study the FRFIM on 
the network.
Solving the ground state problem, we found that it has  the degenerate 
ground states: The black~(white) nodes belong to the $+$~($-$) spin
domain in all ground states, while the gray nodes~(3, 10,
29) may belong to either domain. 
Note that the spin domains almost coincide with the actual shape of the 
network after the breakup; all black~(white) nodes are in the side of the
administrator~(instructor). The gray nodes are in the marginal
state. It is reasonable to think that they do not belong to any
community. In the previous work~\cite{Girvan02},  the
node 3 was misclassified. Our result hints that it is due to the marginality.

The example clearly shows that the FRFIM is useful in
finding out the community structure. 
For general application,
(i) one needs to know the ground state(s) of the FRFIM of Eq.~(\ref{eq:H}) with
the quenched random magnetic field given in Eq.~(\ref{eq:Bst}) for
any node pair of $s$ and $t$. Then, one needs to identify the set of all
nodes that belong to the same spin domain as $s$ and $t$ in {\em
all} ground states. Those sets will be called the {\em cliques} and
denoted by $\mathcal{C}_s$ and $\mathcal{C}_t$, respectively. The
number of nodes in the clique $\mathcal{C}$ will be called the clique size 
and denoted by $|\mathcal{C}|$. (ii) More importantly,
one needs to specify the node pair $s$ and $t$ which is
relevant to the community structure. An arbitrary choice of $s$
and $t$ will not provide any information on the community
structure. For example, if we take $s=12$ and $t=15$ in the
Zachary network in Fig.~\ref{fig:zachari}, we obtain that
$\mathcal{C}_s=\{12\}$ and all other nodes are in $\mathcal{C}_t$.
This merely means that the node 12 is a peripheral node.

For (i), the ground state problem of the FRFIM can be solved exactly
with the help of a numerical combinatorial optimization 
algorithm~(see Appendix). 
This is achieved by mapping the ground state problem 
onto the minimum cut problem or the maximum flow problem~\cite{HeikoRieger}.
The algorithm allows us to find all ground states, with which we can 
find the cliques $\mathcal{C}_s$ and $\mathcal{C}_t$ for any pair of 
$s$ and $t$. We explain the detailed procedure in Appendix.

For (ii), the community structure can be found
from the distribution of the clique sizes for all pairs of $s$ and $t$.
For a certain pair of $s$ and $t$, one may have that
$|\mathcal{C}_s| \sim |\mathcal{C}_t| \sim \mathcal{O}(1)\ll N$.
It happens when $s$ and $t$ are peripheral nodes of the network;
most nodes are not influenced by them.
Such a pair does not provide any information on the community structure.
One may have that
$|\mathcal{C}_s| \sim \mathcal{O}(1) \ll |\mathcal{C}_t| \sim \mathcal{O}(N)$.
This happens when $s$ is a peripheral node while $t$ is inside the bulk.
The cliques $\mathcal{C}_s$ and $\mathcal{C}_t$ do not correspond to 
a community either.
On the contrary, one may have that $\mathcal{O}(1)\ll
|\mathcal{C}_s| \sim |\mathcal{C}_t| \sim \mathcal{O}(N)$.
This happens only when there exist communities whose sizes are of the order
of $N$, $s$ and $t$ are chosen among 
``influential" nodes in different communities.
In this case, we will regard the cliques $\mathcal{C}_s$ and 
$\mathcal{C}_t$ as the communities in the network. 

In order to distinguish the different cases, we define the 
``separability" $D_{st}$ for a node pair $s$ and $t$ 
as the product of the clique sizes,
\begin{equation}\label{eq:SP}
D_{st} = | \mathcal{C}_s | \cdot | \mathcal{C}_t | \ .
\end{equation}
It ranges in the interval $1\leq D_{st} \leq N^2/4$.
We propose that the community structure be detected with the distribution
of the separability $D_{st}$ for all pairs of $s$ and $t$.
If $D_{st} \lesssim O(N)$ for all pairs of $s$ and $t$, then we 
conclude that the network has no community structure.
On the other hand, if $D_{st}\sim O(N^2)$ for a certain pair of $s$ and $t$,
then we conclude that the network consists of communities that can be
identified from the cliques $\mathcal{C}_s$ and $\mathcal{C}_t$.
Moreover, the nodes $s$ and $t$ may be
regarded as influential nodes of the communities.
Therefore, in our method, the existence of the community structure is
verified with the scaling behavior of the maximum value of the separability
with the network size.

For a given network size $N$, the scaling can be examined with the quantity
$\ln D_{st} / \ln N$. 
Without the community structure, it would be close to or much less than 1 
for all node pairs. A node pair with $\ln D_{st} / \ln N>1$ indicates the 
presence of the community structure.

\section{RESULTS}\label{sec:3}
We test the method by applying it to the 
Barab\'asi-Albert~(BA) network~\cite{BAnet},
the Zachary karate club network~\cite{Girvan02}, the scientific
collaboration network~\cite{Girvan02}, and the stock price correlation
network~\cite{DHKim05}. In each network, the separability was calculated 
for all node pairs, and the separability distribution was examined 
with a so-called rank plot, where $[\ln D / \ln N]$ is plotted against a
normalized rank of each node pair. The rank is assigned to each node
pair in the ascending order of the separability. It is then
normalized so that the rank of the pair with the maximum value of
the separability is equal to 1. 

\begin{figure}[t]
\includegraphics*[width=\columnwidth]{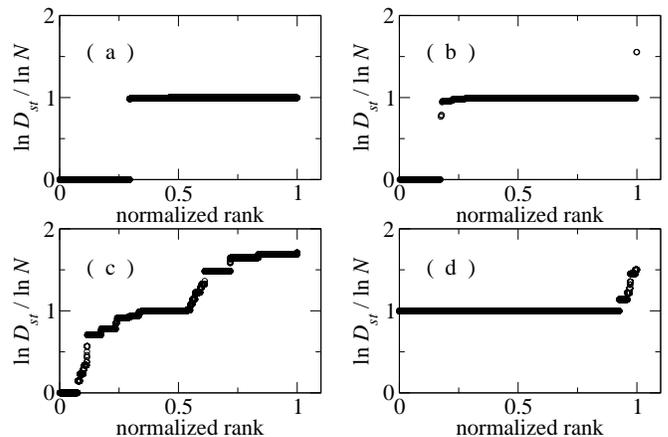}
\caption{The rank plot for the separability distribution for the
BA network (a),  the Zachary karate club network (b), the
scientific collaboration network (c), and the stock price
correlation network (d).} \label{fig:rank_plot}
\end{figure}
\begin{figure}[t]
\includegraphics[width=\columnwidth]{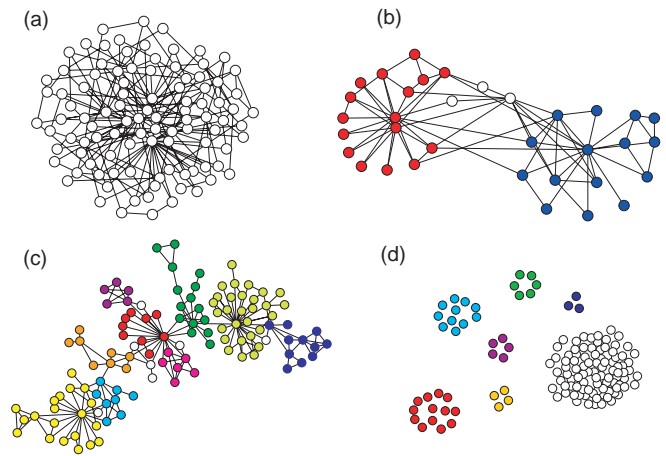}
\caption{(Color online) The community structure of (a) the BA
model network ($N=100$), (b) the Zachary karate club
network~($N=34$), (c) the scientific collaboration
network~($N=118$), and (d) the stock correlation
network~($N=137$). Nodes in different communities are
distinguished with color. The white symbols represent the marginal
nodes.} \label{fig:comm}
\end{figure}

The BA network is an unweighted network. It is known that
the BA network does not have a community structure.
We grew a BA network of $N=100$ nodes,
and calculated the separability $D_{st}$ for all node pairs.
The separability distribution is presented with the rank plot
in Fig.~\ref{fig:rank_plot}~(a).
We find that the separability is clustered at
$D_{st}=1$ and near $D_{st}\simeq N$ for all pairs of $s$ and $t$,
hence $\ln D_{st}/\ln N \lesssim 1$.
This confirms that the BA network does not have the community structure~(see
Fig.~\ref{fig:comm}~(a)), and
demonstrates what the separability distribution looks like for networks
without the community structure.

Next we study the separability distribution of the Zachary karate
club network of $N=34$ nodes, which is presented in
Fig.~\ref{fig:rank_plot}~(b). We found that $D_{st}\lesssim N$ for
all node pairs but $(1,34)$ and $(1,33)$. For the pairs
$(s,t)=(1,33)$ and $(1,34)$, we obtained the same cliques,
$\mathcal{C}_s$ of 15 nodes and $\mathcal{C}_t$ of 16 nodes, which are
marked with the black and the white symbols in
Fig.~\ref{fig:zachari}, respectively~(see also
Fig.~\ref{fig:comm}~(b)). Therefore we can conclude that there
exist two communities in the network and that the node $1$ is an
influential node of one community and the nodes 33 and 34 are of
the other community. In fact the nodes 1 and 34 correspond to the
club instructor and the administrator, respectively. The detected
communities are in good agreement with the network shape after the
breakup.

We also investigate the community structure of a larger and more
complex network. We examine the unweighted collaboration network
of $N=118$ scientists in the Santa Fe Institute~\cite{Girvan02}.
In this network, two nodes~(scientists) are linked if they 
coauthored at least one article. The rank plot
is presented in Fig.~\ref{fig:rank_plot}~(c). One can see that the
separability is distributed broadly, which indicates that the
network has multiple~(more than two) communities.

In such a case, the communities can be identified by applying our method
hierarchically: First of all, one can find the node pair $(s_0,t_0)$ with 
the largest separability, and the corresponding cliques 
$\mathcal{C}_{s_0}$ and $\mathcal{C}_{t_0}$. 
The clique may consist of a single community or
be the union of several sub-communities.
In order to investigate the sub-structure, one constructs the sub-network
which consists of all nodes and links within each clique.
Then, one can apply the method to the sub-networks. 
This can be performed hierarchically until a sub-network
does not have the community structure any more. 
Or one may proceed with the iteration
only when the subnetwork size is equal to or larger than a threshold value $m$.
The resulting cliques can then be identified as communities up to a
resolution $m$.

With the hierarchical application of our method, we find
the community structure of the scientific collaboration network as shown
in Fig.~\ref{fig:comm}~(c). Here, we identify all communities whose
size are equal to or larger than $m=5$. The community structure
is in good agreement with that found in Ref.~\cite{Girvan02}.

Our method is also applicable to weighted networks. As an example of
weighted networks, we study the economic network of 137 companies in the
New York Stock Exchange market.
The network is constructed through the stock price
return correlation between the companies for the 21 year period from 1983 to
2003~\cite{DHKim05}. 
With the stock price $P_i(t)$ of a company $i$ at time $t$, the return
is given by $R_i(t) = \ln P_i (t+\Delta t) - \ln P_i(t)$ with the unit time
interval $\Delta t$ taken to be one day. Then, the stock price correlation
is given by
$$C_{ij} = \frac{\langle (R_i-\langle R_i\rangle )
( R_j-\langle R_j\rangle)\rangle }{ \sqrt{(\langle R_i^2\rangle
-\langle R_i \rangle^2 ) ( \langle R_j^2\rangle -\langle R_j \rangle^2 )}} \
, $$
where the angular bracket indicates the time average over the period.
Its value ranges in the interval $-1\leq C_{ij} \leq 1$, and is large for
strongly correlated company pairs. It has been shown that the structural
information of the economic system is encoded in the correlation
matrix $\{C_{ij}\}$~\cite{Mantegna00,Noh00}.

In order to apply our method,  all weights are required to be
non-negative. Hence, we assume that the weight is given by $J_{ij} =e^{a
C_{ij}}$ with a positive constant $a$ taken to be $20$.
The weights are positive for all pairs of nodes,
and the economic network is fully connected.
The separability distribution is shown in the rank plot in
Fig.~\ref{fig:rank_plot}~(d). As in the collaboration network, there are several
nontrivial separability levels. We identified all communities whose size is
equal to or larger than 3 with the same hierarchical method as in the 
collaboration network. The resulting shape of the network is illustrated in
Fig.~\ref{fig:comm}~(d). We confirmed that the communities are formed by
companies in the same industrial sector. 
For example, the largest community
consists of 13 companies in the energy sector.
This study shows that our method works well for weighted networks.
We note that many nodes~(white symbols) remain unclassified.
We attribute it to the fully-connectedness of the network.

\section{CONCLUSIONS}\label{conclusions}

In this paper we have proposed the method for finding out the
community structure of general networks. It is achieved by
studying the ground state problem of the FRFIM on the networks
with the magnetic field distribution given in Eq.~(\ref{eq:Bst})
for two arbitrary nodes $s$ and $t$. The cliques $\mathcal{C}_s$
and $\mathcal{C}_t$ are defined as the sets of all nodes that
belong to the same spin domains as $s$ and $t$ in all possible
degenerate ground states, respectively. The community structure is
then manifested in the clique pattern for the pair with the
maximum value of the separability $D_{st}$ defined in
Eq.~(\ref{eq:SP}). Our method is motivated from the observation on
the Zachary karate club network, which shows that the resulting
shape of the network after breakup is determined by the underlying
community structure. In our method, the response of the networks
subject to schism is simulated with the FRFIM.

In our method one can verify the existence of the community structure of
a given network with the scaling property of the separability:
If the separability scales as $D_{st}\lesssim\mathcal{O}(N)$ for all node pairs
as in the BA network,
the network does not have the community structure.
On the other hand, if $D_{st}\sim\mathcal{O}(N^2)$ for a certain pair of
$s$ and $t$, one can conclude that the network has the community structure
and that the nodes $s$ and $t$ are influential nodes in each community.
Another advantage of our method is that it can be applied to
both unweighted and weighted networks. Figure~\ref{fig:comm} shows the
performance of the method in real-world networks.

One of the weak points of our method is the time complexity. Practically the
ground state problem of the FRFIM in sparse networks of $N$ nodes has the
time complexity of $\mathcal{O}(N^\theta)$ with $\theta\simeq
1.2$~\cite{HeikoRieger}. Since one has to solve the ground state problems
for all magnetic field distributions, the total time complexity
scales as $\mathcal{O}(N^{2+\theta})$. Hence, in the practical sense,
our method is limited to networks of up to a few thousands of nodes.
One may avoid the time complexity problem if the important nodes are known
{\em a priori}. In the network theory, importance of nodes can be measured
by, e.g., the degree or the betweenness centrality.
Hopefully the community structure of large networks can be studied if one
incorporates such importance measure into our method.

\begin{acknowledgements}
This work was supported by Korea Research Foundation
Grand(KRF-2003-003-C00091).
JDN would like to thank KIAS for the hospitality during the visit.
\end{acknowledgements}

\appendix*
\section{Minimum cut and maximum flow problem}\label{sec:appendix}
This Appendix is intended to introduce the combinatorial optimization
algorithm solving the ground state problem of the FRFIM.
For more rigorous description, we refer the readers to Ref.~\cite{HeikoRieger}.

Consider a network $G$ of $N$ nodes with the symmetric weight matrix
$\{J_{ij}\geq 0\}$~$(i,j=1,\cdots,N)$.
The ferromagnetic random field Ising model on $G$ is defined by the
Hamiltonian in Eq.~(\ref{eq:H}) with the quenched random magnetic field
$\{B_i\}$. The ground state is the spin configuration that has the minimum
energy among all $2^N$ configurations. One might find the ground state by
enumerating all spin configurations, which is obviously time consuming and
inefficient. We will explain the efficient way for solving the ground state
problem.

\begin{figure}
\includegraphics[width=\columnwidth]{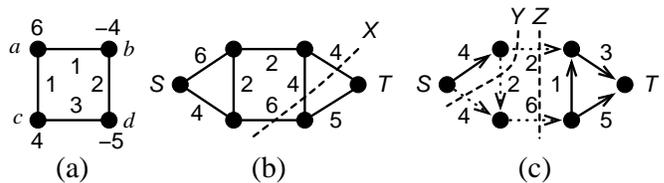}
\caption{
(a) A network $G$ with 4 nodes~(filled circles) and 4 links~(lines) for
the FRFIM.
Figures represent the magnetic fields and the interaction strengths,
respectively.
(b) The corresponding capacitated network $G'$ with the link capacities.
(c) The maximum-flow configuration with $v^\star=8$.
The dotted lines represent saturated links with
$x^\star_{\alpha\beta}=c_{\alpha\beta}$. The dashed lines $X$, $Y$, and $Z$ 
represent boundaries associated with $ST$-cuts.}\label{fig:GandGp}
\end{figure}

It is useful to introduce a capacitated network denoted by $G'$:
Having all nodes and links of $G$,
$G'$ contains two additional nodes $S$, called the source, and
$T$, called the sink, and additional links between the source~(sink)
and the nodes with the positive~(negative) magnetic field.
$G'$ is also a weighted network with the symmetric weight matrix
$\{c_{\alpha\beta}\}$~($\alpha,\beta=S,T,1,\cdots,N$).
For a link $(ij)$ from the original network $G$, the weight is given by
$c_{ij} = 2 J_{ij}$.
For the additional link the weight is given by
$c_{Si} = B_i$ for all $i$ with $B_i>0$ and $c_{iT}=|B_i|$ for all $i$ 
with $B_i<0$.
The weight of the network $G'$ is usually called the capacity.
Figure~\ref{fig:GandGp} illustrates the relation between a network $G$ of four
nodes $\{a,b,c,d\}$ and the corresponding capacitated network $G'$.

In the capacitated network $G'$ we define a $ST$-cut as a decomposition 
of all nodes into two disjoint sets $\mathcal{S}$ and 
$\mathcal{T}$ with $S\in\mathcal{S}$ and $T\in \mathcal{T}$.
It will be denoted by $[\mathcal{S},\mathcal{T}]$. 
For a given $[\mathcal{S},\mathcal{T}]$, some links connect nodes
in the different sets. The set of such links forms the boundary of the cut, 
which is denoted by $(\mathcal{S},\mathcal{T})=
\{(\alpha\beta)|\alpha\in\mathcal{S},\beta\in\mathcal{T}\}$.
The cut capacity $C[\mathcal{S},\mathcal{T}]$ is then defined as the total
sum of the capacity of the boundary links, that is,
\begin{equation}
C[\mathcal{S},\mathcal{T}] =
\sum_{(\alpha\beta)\in(\mathcal{S},\mathcal{T})} c_{\alpha \beta} \ .
\end{equation}
Figure~\ref{fig:GandGp} shows some examples of the cut.
The boundary denoted by $X$ is associated with a cut
$[\{S,a,b,c\},\{T,d\}]$, whose cut capacity is $14$.

There exists one-to-one correspondence between the Ising spin configuration
on the weighted network $G$ and the cut $[\mathcal{S},\mathcal{T}]$ of the
capacitated network $G'$. It is achieved by assigning $\sigma_i=+1(-1)$
for all nodes $i$ in $\mathcal{S(T)}$ and vice versa.
Hence, the sets $\mathcal{S}$ and $\mathcal{T}$ correspond to up and down
spin domains, respectively, and the  boundary $(\mathcal{S},\mathcal{T})$
corresponds to the spin domain wall.
Furthermore, one can
easily verify that the energy $E$ of the FRFIM of a spin configuration
$\{\sigma_i\}$ and the cut capacity $C[\mathcal{S},\mathcal{T}]$ satisfy the
relation
\begin{equation}
E(\{\sigma_i\}) = C[\mathcal{S},\mathcal{T}] + E_0
\end{equation}
where
$E_0 = -\sum_{i,j}J_{ij}/2 - \sum_{i} |B_i|/2$.
Therefore, solving the ground state of the FRFIM on $G$
is equivalent to finding the optimal $ST$-cut on $G'$ whose cut capacity
is minimum. It is called the {\em minimum cut problem}.

The minimum cut problem can be further mapped on to the {\em maximum flow 
problem}:
On the capacitated network $G'$, a flow is to denote a set of flow variables
$\{x_{\alpha\beta}\}$ defined for all links in $G'$ which are subject to a
capacity constraint
\begin{equation}\label{eq:capacity}
0\leq x_{\alpha\beta} \leq c_{\alpha\beta}
\end{equation}
and a mass balance constraint
\begin{equation}\label{eq:balance}
{\sum_\beta}' x_{\alpha\beta} - {\sum_\beta}' x_{\beta\alpha} =
v \delta(\alpha,S) - v \delta(\alpha,T)  .
\end{equation}
Here $\sum'$ means a sum over all adjacent nodes of $\alpha$,
$\delta()$ denotes the Kronecker $\delta$ symbol,
and $v$ is a non-negative parameter.
The mass balance constraint allows us to interpret the flow
$\{x_{\alpha\beta}\}$ as a conserved flux configuration of a, e.g., fluid
which is originated from the source $S$ by the amount of $v$ and
targeted to the sink $T$ through the network $G'$.

Due to the capacity constraint, there exists the upper bound in $v$,
beyond which a flow satisfying Eqs.~(\ref{eq:capacity}) and
(\ref{eq:balance}) does not exist.
Then, the question that arises naturally is to find the maximum value
$v^\star$ and the corresponding flow $\{x^{\star}_{\alpha\beta}\}$
that can be delivered. This is the maximum flow problem.

The celebrated max-flow/min-cut theorem of Ford and Fulkerson~\cite{Ford62}
states that for a given capacitated network $G'$, the maximum flow
$v^\star$ is equal to the minimum cut capacity, that is to say,
\begin{equation}
v^\star = \min_{[\mathcal{S}, \mathcal{T}]} C[\mathcal{S},
\mathcal{T}] \ .
\end{equation}
The rigorous proof of the theorem can be found elsewhere~\cite{HeikoRieger}.
Intuitively the theorem states that the maximum flow is limited by the
bottleneck in the network whose capacity is given by the minimum cut
capacity.

The maximum flow problem can be solved numerically
in a polynomial time with the augmenting path algorithm or the
preflow-push/relabel algorithm~\cite{Ford62,HeikoRieger}.
In the augmenting path algorithm, one repeatedly searches for a path from
$S$ to $T$ via
{\em unsaturated}~($x_{\alpha\beta}<c_{\alpha\beta})$ links and updates
$\{x_{\alpha\beta}\}$ by augmenting flows along the path.
When the augmenting path does not exist any more, the resulting flow
corresponds to the maximum flow configuration. The preflow-push/relabel
algorithm is a more sophisticated and efficient algorithm. 

Once the maximum flow configuration $\{x^\star_{\alpha\beta}\}$ is found,
the minimum cut is constructed easily. Let $\mathcal{S}_S$ be the set of all
nodes of $G'$ that can be {\em reachable} from the source $S$ only through
{\em unsaturated}~($x^\star_{\alpha\beta} < c_{\alpha\beta}$) links.
Trivially, $\mathcal{S}_S$ does not include the sink $T$, since
there does not exist any augmenting path in the maximum flow configuration.
Hence, the set $\mathcal{S}_S$ and its complement
$\overline{\mathcal{S}_S}$ defines a cut
$[\mathcal{S}_S,\overline{\mathcal{S}_S}]$, which is indeed a minimum cut of
$G'$.

One may find the minimum cut alternatively. 
Let $\mathcal{T}_T$ be the set of all
nodes of $G'$ that can be {\em reachable} from the sink $T$ only through
{\em unsaturated} links. Then, $\mathcal{T}_T$ and its complement
$\overline{\mathcal{T}_T}$ defines a cut
$[\overline{\mathcal{T}_T},\mathcal{T}_T]$, which is also the minimum cut.

The two cuts $[\mathcal{S}_S,\overline{\mathcal{S}_S}]$ and
$[\overline{\mathcal{T}_T},\mathcal{T}_T]$ may be different, which implies
that the corresponding FRFIM has degenerate ground states.
In that case, all degenerate ground states can be found
systematically~\cite{HeikoRieger}. In this work, we are interested in
the spins that are fixed in all ground states. One can easily verify that
all nodes $i\in \mathcal{S}_S~(\mathcal{T}_T)$
except for $S~(T)$ are in the spin state $\sigma_i = +1~(-1)$ in all ground
states. The other nodes $j\notin \mathcal{S}_S$ and
$\mathcal{T}_T$ may be in either state $\sigma_j = \pm 1$.

We provide an example illustrating the mapping between the FRFIM and the
maximum flow or the minimum cut problem in Fig.~\ref{fig:GandGp}. 
The maximum flow configuration is depicted in Fig.~\ref{fig:GandGp} (c)
with the maximum flow $v^\star=8$. 
The links drawn with dotted lines are saturated~($x^{\star}_{\alpha\beta}
=c_{\alpha\beta}$). The sets of all nodes that are
reachable from $S$ and $T$ through unsaturated links are given by 
$\mathcal{S}_S=\{S,a\}$ and $\mathcal{T}_T=\{T,b,d\}$. They yield 
the minimum cuts $[\mathcal{S}_S,\overline{\mathcal{S}_S}]$ and
$[\overline{\mathcal{T}_T},\mathcal{T}_T]$ whose boundaries
are $Y$ and $Z$, respectively.
Hence, one finds that $\sigma_{a}=+1$ and $\sigma_{b}=\sigma_{d}=-1$ in all
degenerate ground states. The node $c$ does not belong to neither 
$\mathcal{S}_S$ nor $\mathcal{T}_T$. Hence $\sigma_{c}$ may be either 
$+1$ or $-1$.

In the present work, we consider the FRFIM on a weighted network $G$
with the specific magnetic field distribution
given in Eq.~(\ref{eq:Bst}) for a certain node pair $s$ and $t$. Then, we
need to find the clique $\mathcal{C}_s$~($\mathcal{C}_t$) of $s$~($t$) which
is the set of all nodes that are in the same spin state as
$s$~($t$) in the ground state. We summarize the method to find the cliques:
\begin{enumerate}
\item{ Construct the capacitated network $G'$.}
\item{ Find the maximum flow configuration $\{x^\star_{\alpha\beta}\}$
using the numerical algorithms.}
\item{ Find the set $\mathcal{S}_S$~($\mathcal{T}_T$) of all nodes that
are reachable from $S$~($T$) through unsaturated links with
$x^\star_{\alpha\beta}<c_{\alpha\beta}$.}
\item{ Then, the cliques are given by $\mathcal{C}_s = \mathcal{S}_S -\{S\}$
and $\mathcal{C}_t = \mathcal{T}_T -\{T\}$.}
\end{enumerate}
After finding the cliques, the community structure can be investigated with
the method explained in Sec.~\ref{sec:2}.


\begin{references}

\bibitem{Albert02} R. Albert and A.-L. Barab\'asi, 
                   Rev. Mod. Phys. {\bf 74}, 47 (2002).

\bibitem{Dorogovtsev02} S.N. Dorogovtsev and J.F.F. Mendes, 
                   Adv. Phys. {\bf 51}, 1079 (2002).

\bibitem{Newman03R} M.E.J. Newman, SIAM Rev. {\bf 45}, 167 (2003).

\bibitem{Jeong00} H. Jeong, B. Tombor, R. Albert, Z.N. Oltvai, and
A.-L. Barab\'asi, Nature (London) {\bf 407}, 651 (2000).

\bibitem{Holme03} P. Holme, M. Huss, and H. Jeong, 
                  Bioinformatics {\bf 19}, 532 (2003).

\bibitem{Wilkinson04} D. Wilkinson and B.A. Huberman, 
                  Proc. Natl. Acad. Sci. {\bf 101}, 5241 (2004).

\bibitem{Ravasz03} E. Ravasz, A.L. Somera, D.A. Mongru, Z.N. Oltvai, and
                   A.-L. Barab\'asi, Science {\bf 297}, 1551 (2002); 
        E. Ravasz and A.-L. Barab\'asi, Phys. Rev. E {\bf 67}, 026112 (2003).

\bibitem{Mantegna00} R.N. Mantegna, Eur. Phys. J. B {\bf 11}, 193 (1999) ;
        G. Bonanno, G. Caldarelli, F. Lillo, and R.N. Mantegna, 
        Phys. Rev. E {\bf 68}, 046130 (2003).

\bibitem{Onnela03} J.-P. Onnela, A. Chakraborti, K. Kaski, J.
Kertesz, and A. Kanto, Phys. Rev. E {\bf 68}, 056110 (2003).

\bibitem{Girvan02} M. Girvan and M.E.J. Newman, 
        Proc. Natl. Acad. Sci. {\bf 99}, 7821 (2002).

\bibitem{Newman04a} M.E.J. Newman, Eur. Phys. J. B {\bf 38}, 321 (2004).

\bibitem{Newman04} M.E.J. Newman and M. Girvan, 
                   Phys. Rev. E {\bf 69}, 026113 (2004).

\bibitem{Newman01} M.E.J. Newman, Phys. Rev. E {\bf 64}, 016131 (2001);
                  {\em ibid.} {\bf 64}, 016132 (2001).

\bibitem{Tyler03} J.R. Tyler, D.M. Wilkinson, B.A. Huberman,
                  comd-mat/0303264 (2003).

\bibitem{Radicchi04} F. Radicchi, C. Castellano, F. Cecconi, V.
Loreto, and D. Parisi, Proc. Natl. Acad. Sci. {\bf 101}, 2658 (2004).

\bibitem{Fortunato04} S. Fortunato, V. Latora, and M. Marchiori, 
                      Phys. Rev. E {\bf 70} 056104 (2004).

\bibitem{Newman04b} M.E.J. Newman, Phys. Rev. E {\bf 69}, 066133 (2004).

\bibitem{Clauset04} A. Clauset, M.E.J. Newman, and C. Moore,
                    Phys. Rev. E {\bf 70}, 066111 (2004).

\bibitem{Reichardt04} J. Reichardt and S. Bornholdt,
                      Phys. Rev. Lett. {\bf 93}, 218701 (2004).

\bibitem{Zhou03} H. Zhou, Phys. Rev. E {\bf 67}, 061901 (2003).


\bibitem{Wu03} F. Wu and B.A. Huberman, Eur. Phys. J. B {\bf 38}, 331 (2004).

\bibitem{Newman04c} M.E.J. Newman, Phys. Rev. E {\bf 70}, 056131 (2004).


\bibitem{Middleton02} A.A. Middleton and D.S. Fisher, 
                      Phys. Rev. B {\bf 65}, 134411 (2002).

\bibitem{Noh_Rieger} J.D. Noh and H. Rieger, Phys. Rev. Lett. {\bf 87}, 
        176102 (2001);
        J.D. Noh and H. Rieger, Phys. Rev. E {\bf 66}, 036117 (2002).

\bibitem{Son04} S.-W. Son, H. Jeong, and J.D. Noh, unpublished.

\bibitem{HeikoRieger} M. Alava, P.M. Duxbury, C. Moukarzel, and H. Rieger,
        in {\em Phase Transitions and Critical Phenomena}
        edited by C. Domb and J.L. Lebowitz (Academic, Cambridge, 2000)
        Vol. {\bf 18}, pp. 141-317;
        A. Hartmann and H. Rieger,
        {\it Optimization Algorithms in Physics} (Wiley VCH, Berlin, 2002).

\bibitem{BAnet} A.-L. Barab\'asi and R. Albert, Science {\bf 286}, 509 (1999);
        A.-L. Barab\'asi, R. Albert, and H. Jeong,
        Physica A {\bf 272}, 173 (1999).

\bibitem{DHKim05} D.-H. Kim and H. Jeong, unpublished.

\bibitem{Noh00} J.D. Noh, Phys. Rev. E {\bf 61}, 5981 (2000).

\bibitem{Ford62} L.R. Ford and D.R. Fulkerson, {\em Flows in
Networks}, (Princeton University Press, 1962).


\end{references}
\end{document}